\documentclass{elsart}

\usepackage{epsfig}

\usepackage{amssymb}

\begin{document}

\begin{frontmatter}

\title{Target Structure Independent $^7\vec{\mathrm{Li}}$ Elastic Scattering at Low Momentum Transfers}

\author[label1]{O.A. Momotyuk\thanksref{now1}},
\corauth[cor1]{Corresponding author}
\thanks[now1]{Permanent address: Institute for Nuclear Research, Kyiv, Ukraine.} 
\author[label2]{N. Keeley\corauthref{cor1}},
\ead{keeley@nucmar.physics.fsu.edu}
\author[label1]{K.W. Kemper},
\author[label1]{B.T. Roeder},
\author[label1]{A.M. Crisp},
\author[label1]{W. Cluff},
\author[label1]{B.G. Schmidt},
\author[label1]{M. Wiedeking},
\author[label3]{F. Mar\'echal},
\author[label4]{K. Rusek},
\author[label4]{S.Yu. Mezhevych},
\author[label5]{J. Liendo}
\address[label1]{Department of Physics, Florida State University,
Tallahassee, Florida 32306-4350, USA}
\address[label2]{CEA-Saclay, DSM/DAPNIA/SPhN, F-91191
Gif-sur-Yvette, France}
\address[label3]{Institut de Recherches Subatomiques, CNRS/IN$_2$P$_3$-ULP, F-67037 Strasbourg Cedex 2, France}
\address[label4]{Department of Nuclear Reactions, The Andrzej So\l tan Institute for Nuclear Studies,
Ho\.za 69, PL-00-681 Warsaw, Poland}
\address[label5]{Physics Department, Simon Bolivar University, Caracas, Venezuela}

\date{\today}
\begin{abstract}
Analyzing powers and cross sections for the elastic scattering of polarized $^7$Li by targets of $^6$Li, 
$^7$Li and $^{12}$C are shown to depend only on the properties of the projectile for momentum transfers 
of less than 1.0 fm$^{-1}$. The result of a detailed analysis of the experimental data within the framework
of the coupled channels model with ground state reorientation and transitions to the excited states of the 
projectile and targets included in the coupling schemes are presented. This work suggests that nuclear 
properties of weakly-bound nuclei can be tested by elastic scattering experiments, independent of the 
target used, if data are acquired for momentum transfers less than $\sim$ 1.0 fm$^{-1}$.
\end{abstract}

\end{frontmatter}

It has been known for some time \cite{blair,parks,hnizdo,tungate,rusek,ni58} that the structure of a heavy-ion projectile determines the characteristics of its elastic scattering. For $^{10}$B, a nucleus with a large ground state quadrupole moment, the addition of a quadrupole reorientation term to the optical potential decreases the depth of the minima in the elastic scattering cross section angular distribution at large scattering angles when compared with $^{12}$C scattering from the same target \cite{parks}. The clearest signature of the influence of the projectile's internal structure on the elastic scattering was provided by comparing $^6$Li and $^7$Li vector and tensor analyzing powers (APs) for energies close to the Coulomb barrier \cite{rusek,ni58}. The difference in sign of the vector analyzing powers between $^6$Li and $^7$Li scattering was shown to arise from strong coupling between the ground, bound and resonant excited states of these loosely bound nuclei. This difference in analyzing powers was also shown to be present for nuclear dominated scattering \cite{he4,c12}. The strong influence of the projectile structure as observed through the analyzing powers is made possible by the weak spin-orbit force in the projectile-target interaction \cite{amak76}.

The surprising result to date is that no clear signature for the structure of the target has been observed in Li scattering APs. This could be a result of the limited number of data sets, or the fact that none of the targets are loosely bound. For this reason, we have undertaken a study of $^7$Li scattering from the loosely bound targets $^6$Li and $^7$Li. The $^6$Li nucleus is bound by 1.47 MeV and it possesses a triplet of states ($3^+$, $2^+$, $1^+$) that are strongly excited in nuclear scattering. While $^7$Li is more tightly bound (2.45 MeV), it has a large ground state quadrupole moment that results in a strong reorientation during scattering as well as having strongly coupled excited states. These two targets have different properties that significantly affect their elastic scattering and might yield knowledge about the importance of target choice when extracting the properties of loosely bound nuclei like $^6$He and $^{11}$Li from elastic scattering.

\begin{figure}
\begin{center}
\psfig{figure=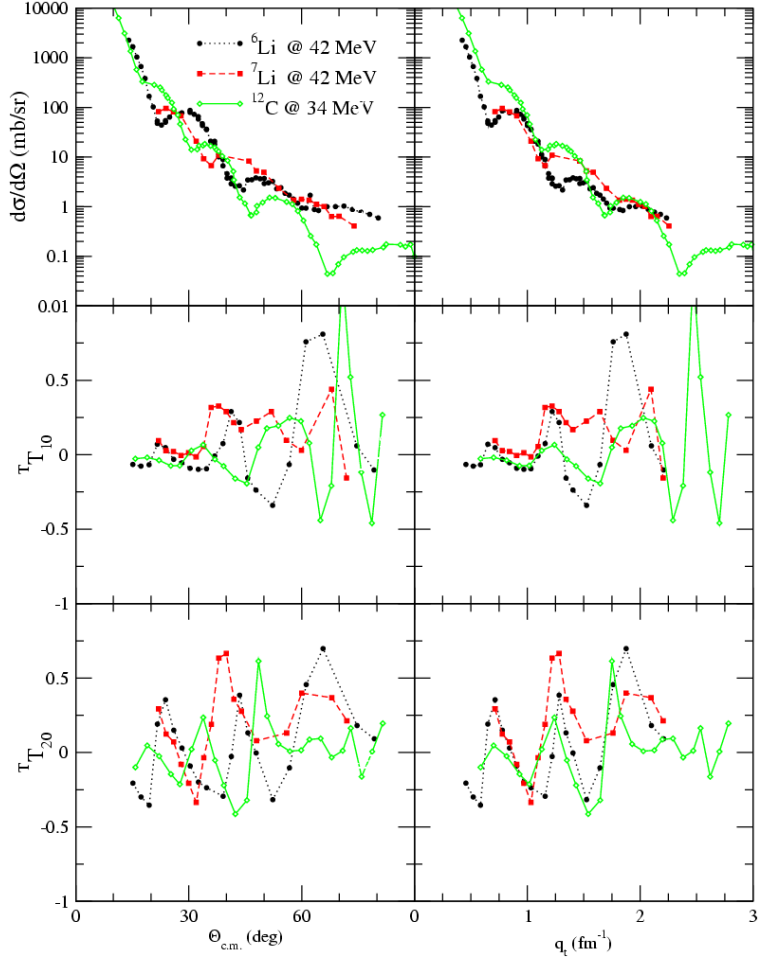,angle=-90,clip=,width=10cm}
\end{center}
\caption{Experimental data for the differential cross section, vector analyzing power $^TT_{10}$ and second rank tensor analyzing power $^TT_{20}$ for $^7\vec{\mathrm{Li}}$ elastic scattering from $^6$Li, $^7$Li and $^{12}$C \cite{c12} plotted versus angle (centre of mass) and the momentum transfer. Error bars are omitted for clarity and the lines are drawn to guide the eye.
\label{fig:1}}
\end{figure}

In this work, we report new cross section and analyzing power angular distributions for the scattering of polarized $^7$Li beams from targets of $^6$Li and $^7$Li. A polarized $^7$Li beam was produced by the Florida State University (FSU) optically pumped lithium ion source \cite{opplis} and accelerated to 42 MeV by the FSU Tandem/Linac accelerator. This energy was chosen because of well known analyzing powers that allow the beam polarization to be determined \cite{he4,gt}. The beam polarizations were: $t_{10} = 0.76 \pm 0.07$ and $t_{20} = 0.37 \pm 0.07$. The targets were enriched in $^6$Li to 95\% and transferred under vacuum to the main scattering chamber. Absolute cross section normalisations were obtained using the target thicknesses determined by comparison of proton elastic scattering measurements with previously accurately determined absolute cross sections \cite{bing71}. The 5\% content of $^7$Li in the targets was enough to allow both sets of measurements to be carried out simultaneously, thus allowing a direct comparison between these two data sets. The left panel in Fig.\ \ref{fig:1} shows the data plotted in the traditional way as a function of angle. 
Also included  are previously reported data for 34 MeV $^7$Li scattering from $^{12}$C \cite{c12}. It is quite difficult to discern any pattern in the displayed data, primarily because the oscillations in the data are determined by the size of  the target nuclei and the energy of the beam relative to the Coulomb barrier, and occur at very different angles. 

It has been found useful in searching for patterns when comparing elastic scattering between the same target and beam but at multiple bombarding energies to display the data as a function of momentum transfer rather than angle \cite{khoa00}, and this was done for the present data sets. The momentum transfer was calculated from $q_t = 2k \cdot \sin (\theta_{\mathrm{c.m.}}/2)$ where $k$ is the wave number of the projectile. 

Fig. \ref{fig:1} shows that the structures of the cross sections and vector analyzing powers are target independent for momentum transfers up to 1.0 fm$^{-1}$, while the second rank tensor analyzing powers $^TT_{20}$ are very similar out to momentum transfer of 1.8 fm$^{-1}$. With this observation, it should be possible through detailed calculations to separate the projectile dependent and projectile-target dependent contributions to the scattering process, and then extract projectile structure information. 

\begin{figure}
\begin{center}
\psfig{figure=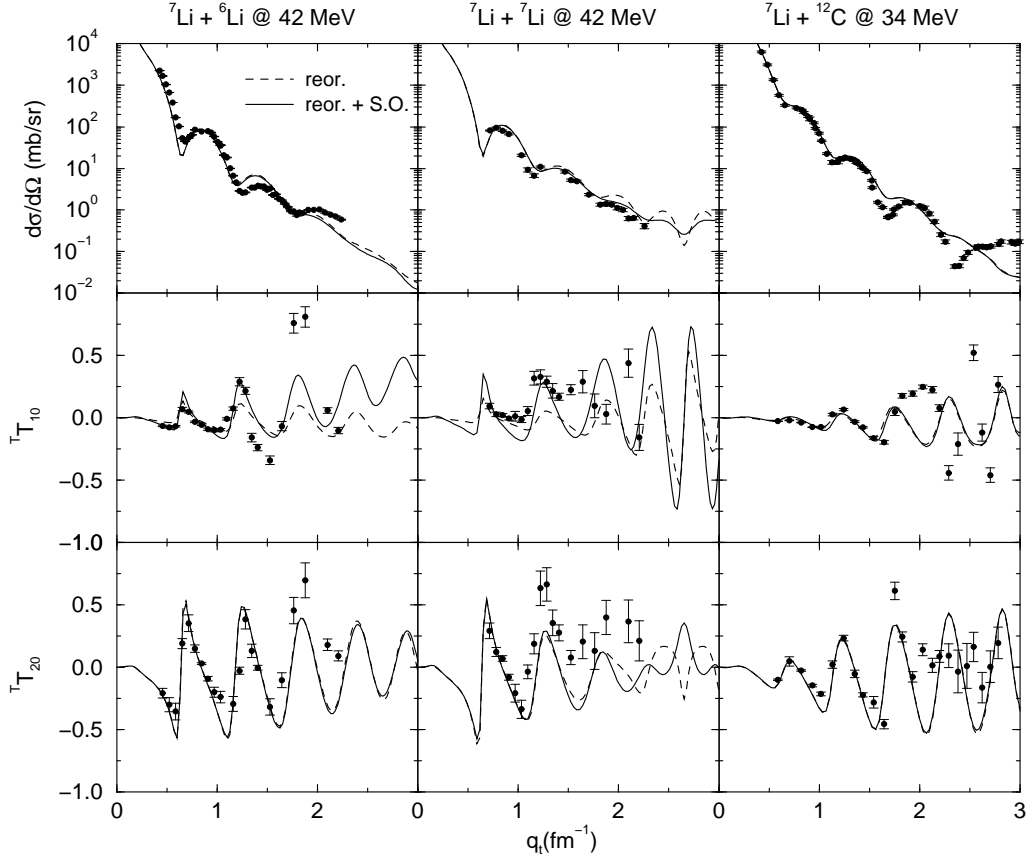,angle=-90,clip=,width=13.5cm}
\end{center}
\caption{Cross section, vector analyzing power $^TT_{10}$ and second rank tensor analyzing power $^TT_{20}$ 
momentum transfer distributions for $^7$Li elastic scattering from $^6$Li and $^7$Li targets at 42 MeV and 
from a $^{12}$C target at 34 MeV. The dashed curves represent the results of CC calculations with coupling 
to the $^7$Li ground state reorientation only. The solid curves denote the results of similar calculations 
including a spin-orbit potential. \label{fig:2}}
\end{figure}

\begin{figure}
\begin{center}
\psfig{figure=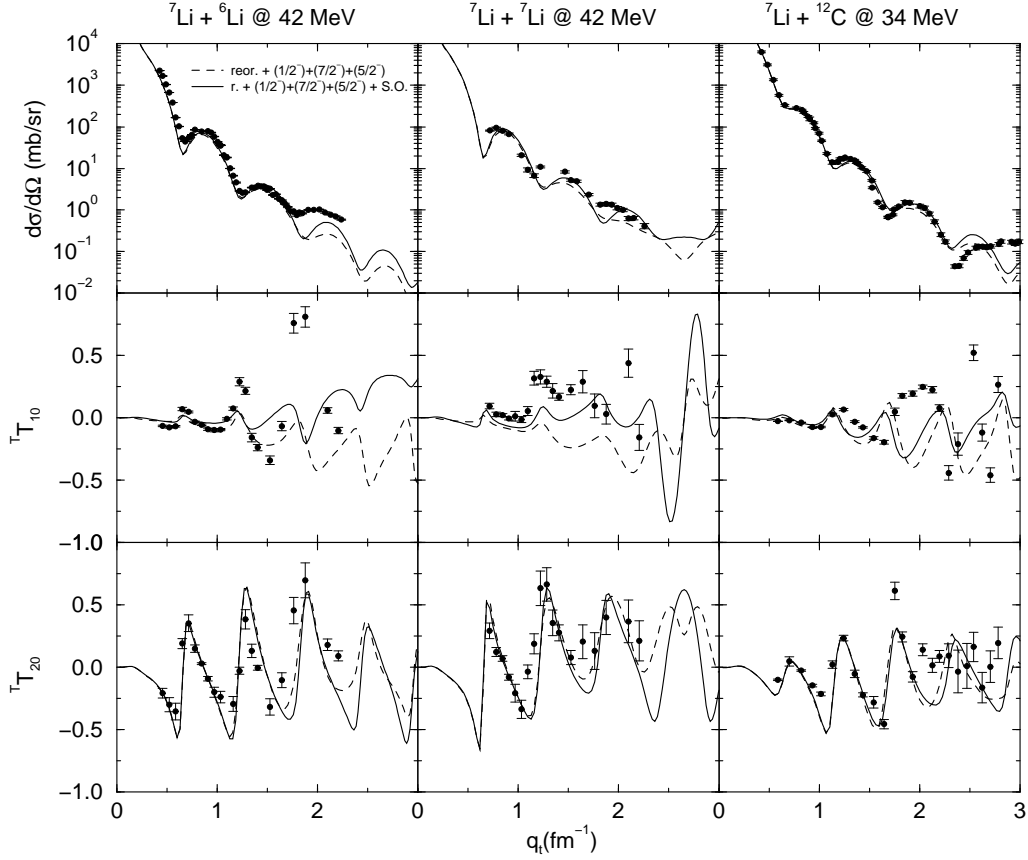,angle=-90,clip=,width=13.5cm}
\end{center}
\caption{Data as in Fig.\ \ref{fig:2}. 
The solid and dashed curves denote the result of the same calculations as in Fig. 2 with added transition to the 3/2-, 1/2- and 7/2- excited states of $^7$Li with and without the real spin-orbit potential, respectively. \label{fig:3}}
\end{figure}

\begin{figure}
\begin{center}
\psfig{figure=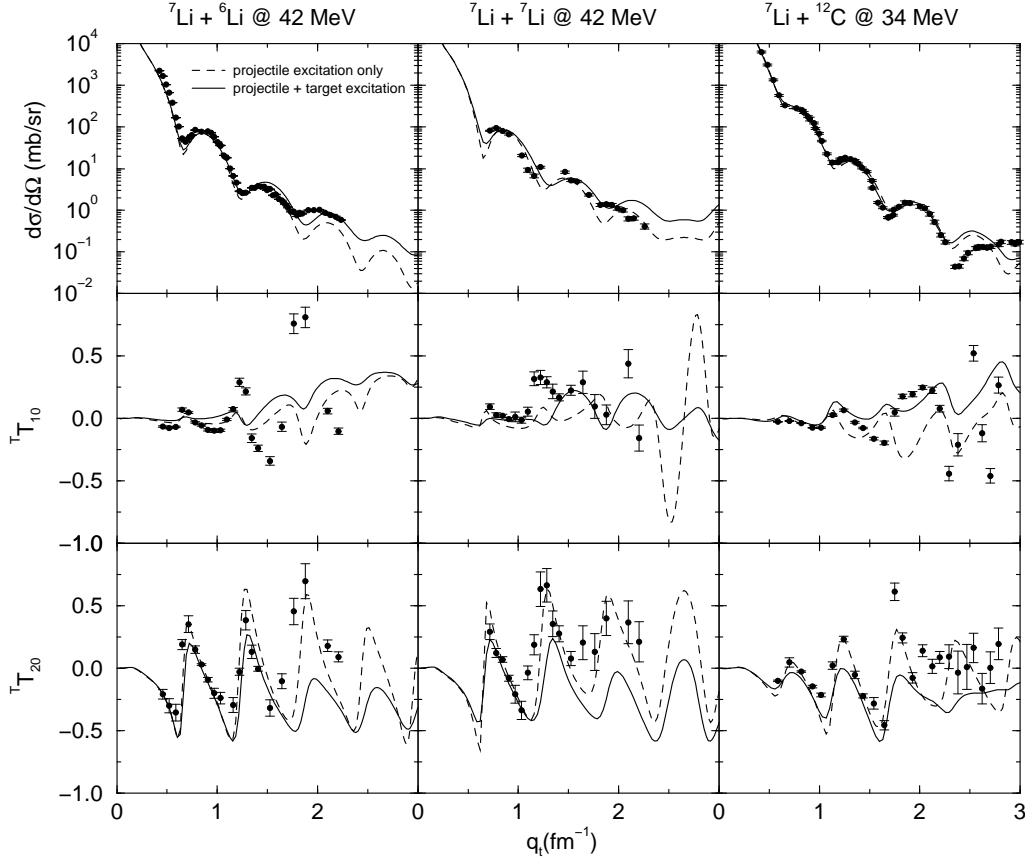,angle=-90,clip=,width=13.5cm}
\end{center}
\caption{Data as in Fig.\ \ref{fig:2}. The curves denote the result of the CC calculations with transition to the 3/2-, 1/2- and 7/2- excited states of $^7$Li couplings only ( dashed line ) as well as added coupling to excited states of targets ( solid line ). \label{fig:4}}
\end{figure}

Such calculations were performed within the framework of the coupled channels (CC) method. Standard optical model (OM) potentials of Woods-Saxon form with volume absorption were used and the parameters are presented in Table \ref{table:1}. 
\begin{table}
\begin{center}
\begin{tabular}{ l c c c c c c c c c }\hline
 & $V$ & $r_V$ & $a_V$ & $W$ & $r_W$ & $a_W$ & $V_{\mathrm{so}}$ & $r_{\mathrm{so}}$ & $a_{\mathrm{so}}$\\ \hline
$^6$Li & 107.8 & 0.750 & 0.853 & 37.9 & 1.000 & 0.757 & 2.5 & 0.950 & 0.45\\
$^7$Li & 107.8 & 0.750 & 0.855 & 37.9 & 0.910 & 0.757 & 2.5 & 0.950 & 0.45\\
$^{12}$C & 107.8 & 0.750 & 0.800 & 47.9 & 1.000 & 0.757 & 1.75 & 0.926 & 0.45\\ \hline
\end{tabular}
\end{center}
\caption{Optical model parameters for $^7$Li elastic scattering from $^6$Li and $^7$Li at 42 MeV and from $^{12}$C at 34 MeV. For all radii the convention $R_i = r_i(A^{1/3}_t + A^{1/3}_p)$ is used.\label{table:1}}
\end{table}
The potential parameters were taken from ref.\ \cite{janecke}, and were slightly modified in order to obtain a better description of the experimental data. The CC calculations were carried out using the code FRESCO \cite{fresco}, version FRXY.1i. The r\^ole of different processes in the $^7$Li interaction was investigated using various coupling schemes. The results of the calculations are shown in Figs.\ \ref{fig:2}-\ref{fig:4}.

The first calculation investigated the r\^ole of the ground state reorientation coupling in the projectile. The standard rotational model was employed with coupling strengths taken from ref.\ \cite{hnizdo}. 
The observed $^TT_{20}$ analyzing powers are reproduced by this process in the forward angular region (dashed curves in Fig.\ \ref{fig:2}) up to $\theta = 60^\circ$ (or $q_t = 1.5$ fm$^{-1}$ momentum transfer) for scattering from $^6$Li and $^{12}$C, and up to $\theta \sim 40^\circ$ ($q_t \sim 1.3$ fm$^{-1}$) for scattering from $^7$Li (in the region of the first two maxima, in agreement with previous work on $^7$Li scattering on $^4$He \cite{he4}, $^{12}$C \cite{c12}, and $^{26}$Mg \cite{mg26}). The projectile ground state reorientation process also partly reproduces the $^TT_{10}$ analyzing powers, providing a rather good description at small momentum transfers, but the amplitudes of the oscillations at the larger momentum transfers are small compared to the data. 

To understand whether the spin-orbit potential gives a significant contribution to the scattering process, further calculations were performed with a real spin-orbit interaction included and the results are also shown in Fig.\ \ref{fig:2} (solid curves). As can be seen, while the inclusion of the spin-orbit potential has essentially no effect on $^TT_{20}$, it significantly increases the magnitude of the oscillations in $^TT_{10}$, particularly at larger momentum transfers, except for the $^{12}$C target where it has little effect. This result is an obvious indication that the spin-orbit interaction, while weak, can still make an important contribution to the scattering process. The lack of effect for the $^{12}$C target is at least partly due to the weaker potential employed for this target -- the use of a stronger potential destroyed the good agreement with $^TT_{10}$ at low momentum transfer generated by the ground state reorientation coupling.

To investigate the r\^ole of coupling to the excited states of $^7$Li (bound $1/2^-$ and $7/2^-$ and $5/2^-$ resonances) as well as target-dependent effects, CC calculations with corresponding coupling schemes were performed and their results are shown in Figs.\ \ref{fig:3}-\ref{fig:4}. Initially, couplings to the excited states of $^7$Li only were included. The rotational model was again employed, and all allowed quadrupole couplings between these states were included. The results are shown in Fig.\ \ref{fig:3} as the solid and dashed curves
for calculations with and without the real spin-orbit potential, respectively. The inclusion of these couplings necessitated a retuning of the optical potential parameters for $^7$Li + $^{12}$C in order to recover reasonable agreement with the cross section; the new parameters were: $V = 107.8$ MeV, $r_V = 0.846$ fm, $a_V = 0.80$ fm, $W = 31.65$ MeV, $r_W = 1.00$ fm, $a_W = 0.757$ fm (the spin-orbit potential parameters were unchanged). 

Adding couplings to the excited states of $^7$Li gives a better description of $^TT_{20}$ for $^7$Li + $^6$Li and $^7$Li + $^7$Li scattering in the region of the second and third maxima while tending to damp out their magnitudes for $^7$Li + $^{12}$C, giving a worse description of this observable. However, they do not play any significant r\^ole in producing the second rank analyzing powers in the momentum transfer region up to 1.5 fm$^{-1}$. Including these processes in the coupling scheme tends to worsen the agreement between the calculated and measured vector analyzing powers, even when the OM potential with the real spin-orbit potential is used (dashed and solid curves in Fig.\ \ref{fig:3}). We again see that the inclusion of a real spin-orbit potential has little or no influence on $^TT_{20}$, its main effect being on $^TT_{10}$ for momentum transfers of greater than about 1.0 fm$^{-1}$. Overall, we may conclude that the effect of couplings to the excited states of $^7$Li on the first and second rank analyzing powers is relatively unimportant compared to that due to ground state reorientation, with the exception of $^TT_{20}$ at momentum transfers greater than about 1.5 fm$^{-1}$ for $^7$Li + $^7$Li scattering.

Couplings to the first excited states of the targets do not play an important r\^ole at small momentum 
transfers in the $^7$Li-target interactions studied here and their inclusion only slightly modifies the 
analyzing powers in the range $0 \leq q_t \leq 1.0$ fm$^{-1}$, as shown in Fig.\ \ref{fig:4} by the solid 
curves. Adding couplings to the excited states of the targets again gives a somewhat worse description of 
the experimental data in comparison to previous calculations. This could be due to the increase in the 
number of exact values which are needed to perform the theoretical calculations. However, the description 
obtained is good enough to define the region in the momentum transfer distribution where the dominant r\^ole 
is played by projectile structure. Couplings for $^7$Li and $^{12}$C used the rotational model, the coupling 
strengths for $^{12}$C being taken from ref.\ \cite{vineyard}, while the couplings for $^6$Li followed the 
method in ref.\ \cite{reber}. The ``target'' excitation coupling for the $^7$Li + $^7$Li system also included ground state reorientation. The $^7$Li + $^{12}$C potentials were again adjusted in order to recover a reasonable description of the cross section data, the new parameters being: $V = 81.92$ MeV, $r_V = 0.850$ fm, $a_V = 0.80$ fm, $W = 14.59$ MeV, $r_W = 1.00$ fm, $a_W = 1.10$ fm (the spin-orbit potential parameters were again unchanged). The large imaginary diffuseness $a_W$ is required if a reasonable description of the data is to be obtained, $\chi^2$ minimizations confirming that this large diffuseness cannot be compensated for by an increase in the imaginary radius $r_W$ when coupling to the $^{12}$C $2^+_1$ state is included.

Including these target excitations in the coupling schemes tends to damp out the oscillations in both the cross sections and analyzing powers for all $^7$Li scattering reactions analyzed here (cf.\ the dashed and solid curves in Fig.\ \ref{fig:4}). 
In general, one may conclude that target excitation effects are only significantly manifest for momentum transfers 
greater than about 1.0 fm$^{-1}$. It should be manifest that while the ground state reorientation coupling for $^7$Li 
as a {\em projectile} is very important, particularly for the generation of the second rank tensor analysing power $^TT_{20}$, when this coupling is included for the $^7$Li {\em target} it has little or no effect. This remarkable result suggests that, at least for small momentum transfers, the target is very much a ``spectator'' as far as the polarisation observables are concerned.

In conclusion, $^7$Li elastic scattering cross section and analyzing power data when graphed as a function of 
momentum transfer rather than angle, show similar structure for momentum transfers less than 1.0 fm$^{-1}$ in 
the nuclear scattering energy regime studied here. While the internal structures of $^6$Li and $^7$Li have been 
shown to strongly influence elastic scattering when they are used as beams, their influence on the scattering 
is not readily apparent when they are used as targets. Although the analyzing powers induced by polarized beams 
of Li have provided a rich environment for investigating the effects of projectile excitation and reorientation, 
a detailed picture of the scattering has not emerged for larger momentum transfers at energies in the nuclear 
scattering regime. The calculations carried out to date do not show any clear dependence on the excitation of 
the target for any system studied. This result suggests that the properties of loosely bound projectiles such 
as $^6$He, $^8$B and $^{11}$Li could be probed by elastic scattering no matter what target is used if the 
momentum transfers are not too large.

This work was supported in part by the State of Florida, the U.S. National Science Foundation and NATO.

\end{document}